\documentclass[12pt]{article}
\title{Phase diagrams of the Spin-1 Blume-Emery-Griffiths Model in a Random Transverse Field}

\author{ H. Ez-Zahraouy$^{(a,*)}$, H. Mahboub$^{(a,b)}$, A.Benyoussef$^{(a)}$, M. J. Ouazzani$^{(b)}$}

\begin{document}
\maketitle
\begin{center}
{\it \small
(a) Facult\'e des Sciences, D\'epartement de Physique, Laboratoire de Magn\'etisme et Physique des Hautes Energies, B.P. 1014, Rabat, Morocco\\
(b) Laboratoire de physique du solide, P.O.Box 1796, Facult\'e des Sciences Dhar Mahraz, F\`es, Morocco}
\end{center}

\begin{center}
{\Large{\bf{	Abstract}}}
\end{center}
The effect of the random quantum transverse field $\Omega$ on the tricritical behavior of the spin-1 Blume- Emery- Griffiths (BEG) model is studied using effective field theory. It is found, that the tricritical behavior depends on both the biquadratic interaction $K$, single- ion anisotropy $\Delta$, and the concentration $p$ of the disorder of $\Omega$. Indeed, there exist a special value $p_{1}$ of the probability $p$ below which the tricritical behavior disappears. In addition, at sufficiently low temperature, the system exhibits long-range order accompanied by the tricritical behavior below a special value $p_{2}$ of the probability $p$.\\ 
\noindent ----------------------------------- \newline
Keys words: Blume-Emery-Griffiths, random transverse field, spin-1, tricritical\\
PACS :05.50.+q; 64.60.Cn; 75.10.Jm; 75.30.Kz \\
($^{*}$) Corresponding author: ezahamid@fsr.ac.ma

\newpage
{\Large{\bf{1- Introduction}}}\\
  The Blume- Emery- Griffiths (BEG) model $[1]$ is a generalization of the  spin- 1 Ising model. It considers both bilinear (J) and biquadratic (K) nearest- neighbor pair interactions, in which a single- ion anisotropy parameter $\Delta$ is included. With vanishing bi-quadratic interactions, the model is known as the Blume-Capel model $[2,3]$. The BEG model has attracted a great deal of attention since it was originally proposed to describe phase separation and superfluid ordering in $^{3}He- ^{4}He $ mixtures. Since its introduction, the BEG model has been extended using a variety of techniques, to solid-liquid-gas systems, multi-component fluid liquid crystal mixtures $[4,5 ]$, magnetic materials $ [6-8 ]$, critical behavior and multi-critical phase diagrams $[9-12 ]$, the re-entrant phenomenon $[13- 20 ]$, and to study metastable and unstable states $[ 21- 23 ]$. Furthermore it is the simplest model that can be used for modelling the behavior of the liquid, solid and vapour phases of a real materials $[24, 25 ]$. The phase diagrams of the BEG model for $0\leq K/J$ have been studied by the mean field approximation $[26- 28 ]$, the position space renormalization  group method $[29]$, the cluster variation method $[30- 31 ]$, series- expansion methods $[32]$, the transfer- matrix method $[33 ]$, the constant coupling approximation $[34 ]$, linear- chain approximation $[35 ]$, and on the Bethe lattice using exact recurring equations $[36 ]$. On the other hand the BEG model with repulsive biquadratic coupling i.e., $ K/J <0 $ was a subject of interest of many authors $[37- 41]$, these studies have shown a variety of interesting features as for example single and double reentrancy regions and ferrimagnetic phases. Keskin and Ekiz $[42 ]$  have also investigated the thermal variations of the BEG model with repulsive biquadratic interactions by using the lowest approximation of the cluster- variation method. \\
The quantum effects on the tricritical behavior of the spin-1 BEG model are very important. However, many works have been done in this context $[43- 48]$.
Particularly, the effect of  random fields on the phase transitions in quantum Ising systems has been a subject of much interest $[48- 56]$, since they show various types of multicritical phenomena.\\  
Our aim in this paper, is to study the effect of the random quantum transverse field on the behavior of phase diagrams of the spin-1 BEG model using the finite cluster approximation $[57- 59]$. It is found, that the tricritical behavior disappears below a concentration $p_{1}$  of the disorder of $\Omega$. 
In addition, at sufficiently low temperature, it is shown that the system exhibits for any value of $\Omega$, a long-range order accompanied by the tricritical behavior, below a special value of the probability $p_{2}$. $p_{1}$ and $p_{2}$ depend on both biquadratic interaction $K$ and single site anisotropy $\Delta$.\\ 
The paper is organized as follows: in section 2 the model and the method are well described; in section 3 results and discussions are given;  section 4 is reserved to the conclusion.\\

{\Large{\bf{2- Model and method}}}\\
The spin-1 BEG model in a transverse field is described by the following Hamiltonian\\

\begin{equation}
H=-J\sum_{(i,j)}S_{iz}S_{jz}-K\sum_{(i,j)}S_{iz}^{2}  S_{jz}^{2}+\Delta\sum_{i}S_{iz}^{2}-\sum_{i}\Omega_iS_{ix}
\end{equation}

Where J and K represent respectively the bilinear and bi-quadratic couplings,  $\Delta $ is the single site anisotropy, $\sum_{(i,j)}$ indicates summation over nearest-neighbor sites of the simple cubic lattice, $\sum_{i}$ indicates summation over all the sites of the lattice, while  $ S_{ix}$ and $ S_{iz}$ are the x and z-components of the spin-1 Pauli matrix respectively given by: \\.

$$ S_{ix}=\frac{\rm 1}{\rm \sqrt{2}}
\pmatrix{0&1&0\cr
         1&0&1\cr
	 0&1&0\cr}
$$
$$ S_{iz}=
\pmatrix{1&0&0\cr
         0&0&0\cr
	 0&0&-1\cr}
$$

$ \Omega_i $ is the transverse magnetic field, which is given by the following distribution law\\
\begin{equation}
P(\Omega_{i})= p\delta(\Omega_{i}- \Omega)+(1-p)\delta(\Omega_{i})
\end{equation}
Where $p$ is the transverse field concentration and $\delta(x)$ is the Dirac function defined as follows:
\begin{equation}
\delta(0) = 1 ; \delta(x\ne 0) = 0  
\end{equation}

Using a single cluster approximation method in which attention is focused on a cluster comprising just a single selected spin labelled 0. And the neighboring spins with which it directly interacts, then the Hamiltonian containing 0 namely $H_{0}$ is given by\\

\begin{equation}
H_{0}= AS_{0z}+(\Delta+C)S_{0z}^{2} -\Omega_0S_{0x}
\end{equation}

Where 
\begin{equation}
A= - J\theta ;
  C= -K\eta
\end{equation}
With
\begin{equation}
 \theta = \sum_{j=1}S_{jz}; 
  \eta = \sum_{j=1}(S_{jz})^{2}
\end{equation}
are operators functions of $S_{jz}$ $(j=1,...,N)$; N is the coordination number.\\
This single site Hamiltonian can readily be diagonalized and its eigenvalues and eigenvectors are easily obtained. The three eigenvectors corresponding to the eigenvalues

\begin{equation}
\lambda_{k}=\frac{\rm 2}{\rm 3}(\Delta + C+\sqrt[3]{\varrho} cos(\varphi_{k}));    k=1, 2, 3
\end{equation}
With

\begin{equation}
\varphi_{k}=\frac{\rm 1}{\rm 3}arc cos(-27q/2\varrho)+2/3(k-1)\pi
\end{equation}
And 

\begin{equation}
\varrho= \frac{\rm {3\sqrt3}}{\rm 2} \sqrt{27 q^{2}+\vert4r^{3}+27q^{2}\vert}
\end{equation}

\begin{equation}
r=-(A^{2}+\Omega^{2})-(\Delta+C)^{2}/3
\end{equation}

\begin{equation}
q=-(\Delta+C)(2A^{2}-\frac{\rm 2}{\rm9}(\Delta+C)^{2}-\Omega^{2})/3
\end{equation}

Are
\begin{equation}
|\psi_{k}>=\alpha_{k}|+>+\beta_{k}|->+\gamma_{k}|0>
\end {equation}

With
\begin{equation}
\alpha_{k}=
{\vert \Omega(\lambda_{k}-\Delta-C+A) \vert \over 2^{1/2}\sqrt {\Omega^{2}((\lambda_{k}-\Delta-C)^{2}+A^{2})+((\lambda_{k}-\Delta-C)^{2}-A^{2})^{2}}}
\end{equation}

\begin{equation}
\beta_{k}=
{\lambda_{k}-\Delta-C-A \over\lambda_{k}-\Delta-C+A} \alpha_{k}
\end{equation}

\begin{equation}
\gamma_{k}=
\sqrt{2}{\lambda_{k}-\Delta-C-A \over\-\Omega} \alpha_{k}
\end{equation}

In a representation in which $S_{0z}$ is diagonal. The starting point of single-site cluster approximation is a set of formal identities of the type

\begin{equation}
<(S_{0\alpha})^{l}>_c=\frac{Tr_{0}(S_{0\alpha})^{l}exp(-\beta H_{0})}{Tr_{0}exp(-\beta H_{0})}
\end {equation}
Where $({S_{0\alpha}})^{l}$ is the $\alpha$- component of the spin operator $S_{0}$ raised to the power l,  $<({S_{0\alpha}})^{l}>_c$ denotes the mean value of $(S_{0\alpha})^{l}$ for a given configuration $c$ of all other spins, i.e when all other spins, $S_{i}(i\ne 0)$ , have fixed values. $Tr_0$ means the trace performed over $S_{0\alpha}$ only.  $\beta= 1/K_BT$; $T$ is the absolute temperature and $K_B$  is the Boltzmann constant.\\

The equations $(16)$ are not exact for an Ising system in a transverse field, they have nevertheless, been accepted as a reasonable starting point in many studies of that system [60].\\
To calculate $<S_{0\alpha}>_c$ and $<(S_{0\alpha})^{2}>_c$, one has to effect the inner traces in eqs. $(16)$ over the states of the spin $0$ and this is the most easily performed using the eigenstates of equation $(12)$. By setting $l=1$ and $l =2$ in eq.(16)  it turns that\\
\begin{equation}
{<S_{0z}>_c}=\frac{\rm \sum_{k=1}^{3}(\alpha_{k}^{2}-\beta_{k}^{2})exp(-\beta \lambda_{k})}
                {\rm\sum_{k=1}^{3}exp(-\beta \lambda_{k}) }
\end{equation}

\begin{equation}
{<S_{0x}>_c}=\sqrt{2}\frac{\rm \sum_{k=1}^{3}(\alpha_{k}+\beta_{k})exp(-\beta \lambda_{k})}
                {\rm\sum_{k=1}^{3}exp(-\beta \lambda_{k}) }
\end{equation}

\begin{equation}
{<(S_{0z})^{2}>_c}=\frac{\rm\sum_{k=1}^{3}(\alpha_{k}^{2}+\beta_{k}^{2})exp(-\beta \lambda_{k})}
                {\rm\sum_{k=1}^{3}exp(-\beta \lambda_{k})}
\end{equation}

\begin{equation}
{<(S_{0x})^{2}>_c}=\frac{\rm\sum_{k=1}^{3}((\alpha_{k}+\beta_{k})^{2}/2+\gamma_{k}^{2})exp(-\beta \lambda_{k})}
                {\rm\sum_{k=1}^{3}exp(-\beta \lambda_{k})}
\end{equation} 

The magnetizations $m_{\alpha}(\alpha = z,x)$ and the quadrupolar moments $q_{\alpha}(\alpha = z,x)$ are given respectively by

\begin{equation}
m_{\alpha} = <<f_\alpha(\theta, \eta, \Omega_0)>_D>_{spins}
\end{equation}
\begin{equation}
q_{\alpha} = <<g_\alpha(\theta, \eta, \Omega_0)>_D>_{spins}
\end{equation}
With 
\begin{equation}
f_{\alpha}(\theta, \eta, \Omega_0) = <S_{0\alpha}>_c
\end{equation}

\begin{equation}
g_\alpha(\theta, \eta, \Omega_0) = <(S_{0\alpha})^{2}>_c
\end{equation}
 Where $<...>_{spins}$ means the average over all configurations of the nearest neighbors spins $S_{i}(i\ne 0)$ of the site $0$ , and $<...>_D$ denotes the average over all configurations of the disorder of the transverse field. Using the distribution of $\Omega_i$ mentioned above, the average over the disorder of $\Omega_{i}$ of  $f_{\alpha}$ and  $g_{\alpha}$ are given by

\begin{equation}
<f_{\alpha}(\theta, \eta, \Omega_0>_D=   \int f_{\alpha}(\theta, \eta, \Omega_0)P(\Omega_0)d\Omega_0
\end{equation}

\begin{equation}
<g_{\alpha}(\theta, \eta, \Omega_0)>_D=   \int g_\alpha(\theta, \eta, \Omega_0) P(\Omega_0)d\Omega_0 
\end{equation}

However to calculate $m_{\alpha}$ and  $q_{\alpha}$ we use an expansion technique of the spin- 1 cluster identities [59 ].\\
Suppose one considers the general product $\rm\prod_{i=1}^{N}\rm\sum_{t=0}^{2} S_{iz}^{t}$ that contains $3^{N}$ terms. From these terms one may collect together all those terms containing u factors of $S_{iz}^{2}$ and v factors of  $S_{iz}$. Such a group is denoted by $\{ S_{z}^{2}, S_{z}\}_{N, u, v}$. For example, if $N = 3$, $u = 1$ and $v = 2$, then \\
 
\begin{equation}
\{ S_{z}^{2}, S_{z}\}_{3, 1, 2} = S_{1z}^{2}S_{2z}S_{3z} + S_{2z}^{2}S_{1z}S_{3z}+ S_{3z}^{2}S_{1z}S_{2z}    
\end{equation}
Our aim now, is to expand $f_\alpha(\theta, \eta, \Omega_0) $ and 
$g_\alpha(\theta, \eta, \Omega_0)$ in terms of these $\{ S_{z}^{2}, S_{z}\}_{N, u, v}$. Thus if one writes 
\begin{equation}
f_\alpha(\theta, \eta, \Omega_0)= \rm\sum_{v=0}^{N}\rm\sum_{u=0}^{N-v}A_{u, v}^{\alpha}(N,\Omega_0)\{S^2, S \}_{N, u, v} 
\end{equation}

\begin{equation}
g_\alpha(\theta, \eta, \Omega_0)= \rm\sum_{v=0}^{N}\rm\sum_{u=0}^{N-v}B_{u, v}^{\alpha}(N,\Omega_0)\{S^2, S\}_{N, u, v}
\end{equation}
The problem is to find the coefficients $A_{u, v}^{\alpha}(N,Omega_0)$ and $B_{u, v}^{\alpha}(N,Omega_0)$. To achieve this is advantageous to transform the spin-1 system to a spin-1/2 representation containing the Pauli operators $\sigma_{iz}= 1, -1$. This may be accomplished by setting $S_{iz} = \tau_{iz}\sigma_{iz}$ with  $\tau_{iz}=0, 1$. In this representation, $f_\alpha$ and $g_\alpha$ become
\begin{equation}
f_{\alpha}(\theta, \eta, \Omega_0)= \rm\sum_{v=0}^{N}\rm\sum_{u=0}^{N-v}A_{u, v}^{\alpha}(N,\Omega_0)\{\tau_{z}, \tau_{z}\sigma_{z} \}_{N, u, v} 
\end{equation}

\begin{equation}
g_{\alpha}(\theta, \eta, \Omega_0)= \rm\sum_{v=0}^{N}\rm\sum_{u=0}^{N-v}B_{u, v}^{\alpha}(N,\Omega_0)\{\tau_{z}, \tau_{z}\sigma_{z} \}_{N, u, v} 
\end{equation}
And must hold for arbitrary choices of $\tau_{iz}$. Suppose one now chooses the first r out of the N operators $\tau_{iz}$ to be unity, and the remainder zero. Then $(30)$ and $(31)$ become

\begin{equation}
f_{\alpha}(\theta, \eta, \Omega_0)= \rm\sum_{v=0}^{r}\rm\sum_{u=0}^{r-v}A_{u, v}^{\alpha}(N,\Omega_0) C_{u}^{r-v}\{\sigma_{z} \}_{r, v} 
\end{equation}

\begin{equation}
g_{\alpha}(\theta, \eta, \Omega_0)= \rm\sum_{v=0}^{r}\rm\sum_{u=0}^{r-v}B_{u, v}^{\alpha}(N,\Omega_0) C_{u}^{r-v}\{\sigma_{z} \}_{r, v} 
\end{equation}
Where $\{\sigma_{z} \}_{r, v}$ is the sum of all possible products of v spin operators, $\sigma_{iz}$, out of a maximum of r, and the $C_{n}^{m}$ are the binomial coefficients $m! /  n! (m-n)!$. That's,
\begin{equation}
f_{\alpha}(\theta, \eta, \Omega_0)= \rm\sum_{v=0}^{r}b_{v}^{\alpha}(r,\Omega_0)\{\sigma_{z} \}_{r, v} 
\end{equation}
\begin{equation}
g_{\alpha}(\theta, \eta, \Omega_0)= \rm\sum_{v=0}^{r}d_{v}^{\alpha}(r,\Omega_0) \{\sigma_{z} \}_{r, v} 
\end{equation}
With
\begin{equation}
b_{v}^{\alpha}(r,\Omega_0) = \rm\sum_{u=0}^{r-v}A_{u, v}^{\alpha}(N,\Omega_0) C_{u}^{r-v}
\end{equation}
\begin{equation}
d_{v}^{\alpha}(r,\Omega_0) = \rm\sum_{u=0}^{r-v}B_{u, v}^{\alpha}(N,\Omega_0) C_{u}^{r-v}
\end{equation}
The spin-1 problem of eqs. $(28)$ and $(29)$ containing N spins has thus been transformed to a spin-1/2 problem containing r spins $[59]$. The advantage of doing this is that now enables one to use directly the results already established in [61] for the spin 1/2 system.
Specializing the results of [61] to a single group of r spins, one has for the current problem
\begin{equation}
b_{v}^{\alpha}(r,\Omega_0) = 1/ (2^{r}{C_v}^{r})\rm\sum_{i=0}^{r}C_{i}^{r}e_{i}(r,v) f_{i \alpha}(r,\Omega_0)
\end{equation}
\begin{equation}
d_{v}^{\alpha}(r,\Omega_0) = 1/ (2^{r}{C_v}^{r})\rm\sum_{i=0}^{r}C_{i}^{r}e_{i}(r,v) g_{i \alpha}(r,\Omega_0)
\end{equation}
Where
\begin{equation}
e_{i}(r,v)= \rm\sum_{\mu = 0}^{i}(-1)^\mu C_{\mu}^{i} C_{v-\mu}^{r-i}
\end{equation}
And
\begin{equation}
f_{i\alpha}(r,\Omega_0)= f_{\alpha}(r-2i,\Omega_0)
\end{equation}
\begin{equation}
g_{i\alpha}(r,\Omega_0)= g_{\alpha}(r-2i,\Omega_0)
\end{equation}
Once the coefficients $b_{v}^{\alpha}(r,\Omega_0)$ and  $d_{v}^{\alpha}(r,\Omega_0)$ have been calculated, the coefficients $A_{u, v}^{\alpha}(N,\Omega_0)$ and $B_{u, v}^{\alpha}(N,\Omega_0)$ may be found by the following procedure. First, $A_{0, v}^{\alpha}(N,\Omega_0)$ and $B_{0, v}^{\alpha}(N,\Omega_0)$  are got by setting $r = v$ in $(36)$ and $(37)$. That's,
\begin{equation}
A_{0v}^{\alpha}(N,\Omega_0) = b_{v}^{\alpha}(v,\Omega_0) 
\end{equation}
And
\begin{equation}
B_{0v}^{\alpha}(N,\Omega_0) = d_{v}^{\alpha}(v,\Omega_0) 
\end{equation}

Then, the other coefficients $A_{u, v}^{\alpha}(N,\Omega_0)$ and $B_{u, v}^{\alpha}(N,\Omega_0)$ may be obtained by expressing $(38)$ and $(39)$ as a reccurence relation, namely

\begin{equation}
A_{r-v, v}^{\alpha}(\Omega_0)= b_ v^{\alpha}(r, \Omega_0)- \rm\sum_{u=0}^{r-v-1}A_{u, v}^{\alpha}(\Omega_0)C_u^{r-v}
\end{equation}

\begin{equation}
B_{r-v, v}^{\alpha}(\Omega_0)= d_ v^{(\alpha)}(r, \Omega_0)- \rm\sum_{u=0}^{r-v-1}B_{u, v}^{\alpha}(\Omega_0)C_u^{r-v}
\end{equation}
Then the magnetizations $m_\alpha$ and the quadrupolar moments $q_\alpha$, where $\alpha = z, x$,  for an arbitray coordination number $N$, are given by
\begin{equation}
m_{\alpha} = \rm\sum_{v=0}^{N}\rm\sum_{u=0}^{N-v}<A_{uv}^{\alpha}(\Omega_0)>_D<\{S_{z}^{2}, S_z\}_{N, u, v}>
\end{equation}
\begin{equation}
q_{\alpha} = \rm\sum_{v=0}^{N}\rm\sum_{u=0}^{N-v}<B_{uv}^{\alpha}(\Omega_0)>_D<\{S_{z}^{2}, S_z\}_{N, u, v}>
\end{equation}
Using the simplest approximation of the Zernike decoupling of the type\\
$<S_{iz} S_{jz}.............. S_{kz}....> = < S_{iz}>< S_{jz}>...........< S_{kz}>...$ for $i\ne j\ne k\ne...$\\
And seeing that the number of elements of the group $\{{S_z}^{2}, S_z\}_{N, u, v}$ is equal to ${C_u}^{N}{C_v}^{N-u}$, then eqs. $(47)$ and $(48)$ become
\begin{equation}
m_\alpha(\Omega) = \rm\sum_{v=0}^{N}\rm\sum_{u=0}^{N-v}<A_{uv}^{\alpha}(\Omega_0)>_D{m_z}^{v}{q_z}^{u}{C_u}^{N}{C_v}^{N-u}
\end{equation}
\begin{equation}
q_\alpha(\Omega) = \rm\sum_{v=0}^{N}\rm\sum_{u=0}^{N-v}<B_{uv}^{\alpha}(\Omega_0)>_D{m_z}^{v}{q_z}^{u}{C_u}^{N}{C_v}^{N-u}
\end{equation}
With
\begin{equation}
<A_{uv}^{\alpha}(\Omega_0)>_D=pA_{uv}^{\alpha}(\Omega)+(1-p)A_{uv}^{\alpha}(0)
\end{equation}
\begin{equation}
<B_{uv}^{\alpha}(\Omega_0)>_D=pB_{uv}^{\alpha}(\Omega)+(1-p)B_{uv}^{\alpha}(0)
\end{equation}

A first order transition is characterized by the gap of the longitudinal magnetization $m_z$ at the transition temperature. If the magnetization vanishes continuously at the transition temperature, this is the second order transition.\\\\
{\Large{\bf{3- Results and discussions}}}

In the following, the ordered phase is characterized by a non zero longitudinal magnetization $m_z$, while the disordered phase is characterized by $m_z=0$. We denote here after by $T$, $\Omega$, $\Delta$ and $K$ the reduced parameters $T/J$, $\Omega/J$, $\Delta/J$ and $K/J$ respectively.\\
At finite temperature, eqs. $(21)$, $(22)$ are solved numerically for $N=6$ (the simple cubic lattice) using iterative method. By varying the value of the biquadratic coupling $K$, we have distinguished three topologies of phase diagrams in the $(\Omega- T)$ plane for a fixed value of $\Delta = 3.5$, which are, the existence of tricritical behavior, only the second order transition exists, and only first order transition is present. Hence, in the case $K = 0.5$, Fig.1a shows the tricritical behavior which depends also on the probability of the disorder of the transverse field, that's, it exists a critical value $p_{1}$ below which the tricritical behavior disappears, this value depends obviously on the biquadratic coupling. Fig. 1b shows the phase diagram $(\Omega- T)$, where the tricritical behavior persists. So We can say that by fixing the biquadratic coupling value $K$, the probability may determine the transition order, as shown in the line corresponding to $p= 0.2$ in Fig. 1a, in which just the second order transition appears, while there is no first order transition.\\
One feature should attract our attention from these two figures, there is the fact that the first order transition is more dominant in the Fig.1b. In principle, we can say that decreasing biquadratic coupling favors the first order transition. In fact, Fig. 1c shows the phase diagram $(\Omega- T)$ for $K = 0.2$, in which only the first order transition appears for any probability $p$. In contrast, for $K=3$ as is seen in Fig. 1d, the $(\Omega- T)$ gives rise to just the second order transition. More precisely, there exist a special value of the biquadratic interaction $K_{1}$ ($=0.27$ for $p=0.4$) below which there is no tricritical behavior (only first order transition can take place), and when $K$ increases $T_{tri}$ decreases, up to a value of $K=K_{2}$ ($= 0.72$ for $p=0.4$) above which the tricritical behavior disappears (only second order transition can give occurrence), see Fig. 2a. In contrast, by fixing $K$ and increasing $p$, $T_{tri}$ increases, see Fig. 2b. Such that, there exist a special value $p_{1}$ ( $= 2.252$ for $K=0.5$) of the concentration  of the disorder of $\Omega$,  below which the tricritical behavior disappears.\\
 The behavior of the tricritical transverse field as a function of $K$ and $p$ is given in Figs. 3a and 3b respectively. It is found that $\Omega_{tri}$ increases when increasing $K$ (Fig. 3a), while it decreases with $p$ as shown in Fig.3b.\\
Besides, for any value of $\Omega$ and at sufficiently low temperature, the system exhibits a long-range order accompanied with a tricritical behavior below a special value of the probability $p_{2}$. \\
 
{\Large{\bf{	4- conclusion}}}\\

In this work we have investigated the effect of the random transverse field $\Omega$ on the tricritical behavior of the quantum spin-1 Blume- Emery- Griffiths (BEG) model. We have found in the ($\Omega$, $T$) phase diagrams, that, the tricritical behavior depends on both the biquadratic interaction $K$, and the concentration of the disorder $p$ of $\Omega$. It is also found, that, at sufficiently low temperatures, the phase diagrams exhibit long-range order accompanied by the tricritical behavior.

\newpage
\section*{References}
\begin{description}
\item[][1] M. Blume, V.J. Emery and R.B. Griffiths, Phys. Rev. A {\bf4} (1971) 1071
\item[][2] M. Blume, Phys. Rev. {\bf141} (1966) 517.
\item[][3] H.W. Capel, Physica  {\bf32} (1966) 966; {\bf33} (1967) 295;{\bf37} (1967) 423.
\item[][4]J. Lajzerowicz, J. Sivardire, Phys. Rev. A {\bf11} (1975) 2079.
\item[][5] J. Sivardire, J. Lajzerowicz,  Phys. Rev. A {\bf11} (1975) 2090.
\item[][6] A. Hinterman, F. Rys, Helv. Phys. Acta  {\bf 42} (1969) 608.
\item[][7] J. Bernasconi, F. Rys, Phys. Rev. B {\bf4}  (1971) 3045.	
\item[][8] H. H. Chen, P. M. Levy, Phys. Rev. B {\bf7}  (1973) 4267.
\item[][9] A. N. Berker, M. Wortis, Phys. Rev. B {\bf14}  (1976) 4946.
\item[][10] R. R. Netz, A. N. Berker, Phys. Rev. B {\bf47}  (1993) 1519.
\item[][11] A. Falikov, A. N. Berker, Phys. Rev. Lett.  {\bf76}  (1996) 4380.
\item[][12] A. Maritan, M. Cieplak, M. R. Swift, F. Toigo, Phys. Rev. Lett . {\bf69}  (1992) 221.
\item[][13] O. F. De alkantara Bonfim, C. H. Obcemea, Z. Phys. B: Condens. Matter {\bf64}(1986) 469.
\item[][14] T. Kaneyoshi, J. Phys. Soc. Jpn. {\bf56} (1987) 4199.
\item[][15] K. G. Chakraborty, J. Phys. C {\bf21} (1988) 2911.
\item[][16] J. W. Tucker, J. Magn. Magn.  Mater. {\bf80} (1989) 203.
\item[][17] I. P. Fittipalde, T. Kaneyoshi, J. Phys: Condens. Matter {\bf1} (1989) 6513.
\item[][18] J. W. Tucker, J. Appl. Phys. {\bf69} (1991) 6164.
\item[][19] J. W. Tucker, J. Magn. Magn. Mater. {\bf191}  (1992) 104.
\item[][20] R. R. Netz, Europhys. Lett. {\bf17} (1992) 373.
\item[][21] A Rosengren, S. Lapinskas, Phys. Rev. B {\bf47} (1993) 2643.
\item[][22] S. Lapinskas, A Rosengren, , Phys. Rev. B {\bf49} (1994) 15190.
\item[][23] G. Grigelionis, A Rosengren, Physica A {\bf208} (1994) 287.
\item[][24] Lev. D. Gelb, Phys. Rev. B {\bf50}  (1994) 11 146.
\item[][25] A. Benyoussef, H. Ez. Zahraouy, H. Mahboub, M. J. Ouazzani, Physica A {\bf326}    (2003) 220. 
\item[][26] D. Mukamel and M. Blume, Phys. Rev. A  {\bf10}  (1974) 610.
\item[][27] J. Lajzerowicz and J. Sivardière. Phys. Rev. A  {\bf11}  (1975) 2079.
\item[][28] J. Sivardière and J. Lajzerowicz, ibid. {\bf11}  (1975) 2090.
\item[][29] A. N. Berker and M. Wortis, Phys. Rev. B {\bf14}  (1976) 4946.
\item[][30] J. W. Tucker, J. Magn. Magn. Mater. {\bf71} (1987) 27 .
\item[][31] C. Buzano and A. Pelizzola, Physica A {\bf195} (1993) 197 .
\item[][32] D. M. Saul, M. Wortis, and D. Stauffer, Phys. Rev. B {\bf9} (1974) 4964.
\item[][33] Z. Koza, C. Jasuukiewicz, and A. Pekalski, Physica A {\bf164} (1990)191 .
\item[][34] K. Takahashi and M. Tanaka, J. Phys. Soc. Jpn. {\bf46}(1979)1428; {\bf48} (1980)1423 .
\item[][35] E. Albayrak and M. Keskin, J. Magn. Magn. Mater. {\bf203} (2000) 201 .
\item[][36] K. G. Chakraborty and  J. W. Tucker, J.  Magn. Magn. Mater. {\bf54} (1986)1349 .
\item[][37] H. H. Chen and P. M. Levy, Phys. Rev. B. {\bf7} (1973) 4267 .
\item[][38] W. Hoston and A. N. Berker, Phys. Rev. Lett. {\bf67}  (1991) 1027 .
\item[][39] W. Hoston and A. N. Berker, J. Appl. Phys.{\bf70} (1991) 6101.
\item[][40] R. R. Netz, Europhys. Lett. {\bf17} (1992)373.
\item[][41] A. Z. Akheyan and N. S. Ananikian, J. Phys. A{\bf 29}(1996) 721.
\item[][42] C. Ekiz and M. Keskin, Phys. Rev. B. {\bf66} (2002)054105.
\item[][43] P. G. De Gennes, Solid State Commun. {\bf1}(1963) 132.
\item[][44] E. Albayrak and M. Keskin, J. Magn. Magn. Mater. {\bf206} (1999) 83.
\item[][45] H. Ez-Zahraouy, ICTP preprint /93/298
\item[][46] H. Ez-Zahraouy, Physica Scripta. {\bf51} (1995) 310 
\item[][47] Hua Wu, Qing Jiang, and Wen Zhong Shen, Phys. Rev. B {\bf69} (2004) 014104.
\item[][48] C\`ecile Monthus, Phys. Rev. B {\bf69} (2004) 054431. 
\item[][49] S. Galam and A. Aharony, J. Phys. C {\bf14}(1981) 3603; {\bf13} (1980)1065.
\item[][50] V. K. Saxena, Phys. Letters A {\bf90}, 71 (1982). 
\item[][51] A. Benyoussef, H. Ez-Zahraouy, M. Saber, Physica A {\bf198} (1993) 593.
\item[][52] A. Benyoussef, H. Ez-Zahraouy , Phys. Stat. Sol. (b) {\bf179}, (1993) 521
\item[][53] A. Benyoussef, H. Ez-Zahraouy , J. Phys: Condens. Matter {\bf6} (1994)3411.
\item[][54] F. Igloi, R. Juhasz, P. Lajko, Phys. Rev. Lett {\bf86} (2001)1343
\item[][55] A. Dhar and A. P. Young, Phys. Rev. B {\bf68} (2003)134441
\item[][56] G. Refael and D. S. Fisher, Phys. Rev. B  {\bf70} (2004)064409.
\item[][57] N. Boccara, Phys. Letters A {\bf94}  (1983) 185 .
\item[][58] A. Benyoussef and N. Boccara, J. Phys. C {\bf16} (1983)1143.
\item[][59] H. Ez- Zahraouy, M. Saber and J. W. Tucker, J. Magn. Magn. Mater. {\bf118} (1993)129.
\item[][60] F. C. Sa Barreto, I. P. Fittipaldi and B. Zeks. Ferroelectrics {\bf39} (1981)1103.
\item[][61] P. Tomczak, E. F. Sarmento, A. F. Siqueira,and A.R. Ferchmin, Phys. Stat. Sol.(b) {\bf142},551 (1987) 
\end{description}

\newpage
\section*{Figure captions}

\begin{description}
\item[Fig.1] 
       Phase diagrams in the $(\Omega, T)$ plane for $\Delta = 3.5$\\ 
  (a): $K = 0.5$\\
  (b): $K = 0.3$\\
  (c): $K = 0.2$\\
  (d): $K = 3.$\\
\item[Fig.2] 
  (a): The behavior of the tricritical temperature as a function of $K$ for $p = 0.4$\\ 
  (b): The behavior of the tricritical temperature as a function of $p$ for $K = 0.5$\\
\item[Fig.3] 
  (a): The behavior of the tricritical transverse field as a function of $K$ for $p = 0.4$\\ 
  (b): The behavior of the tricritical transverse field as a function of $p$ for $K = 0.5$\\ 
  \end{description}
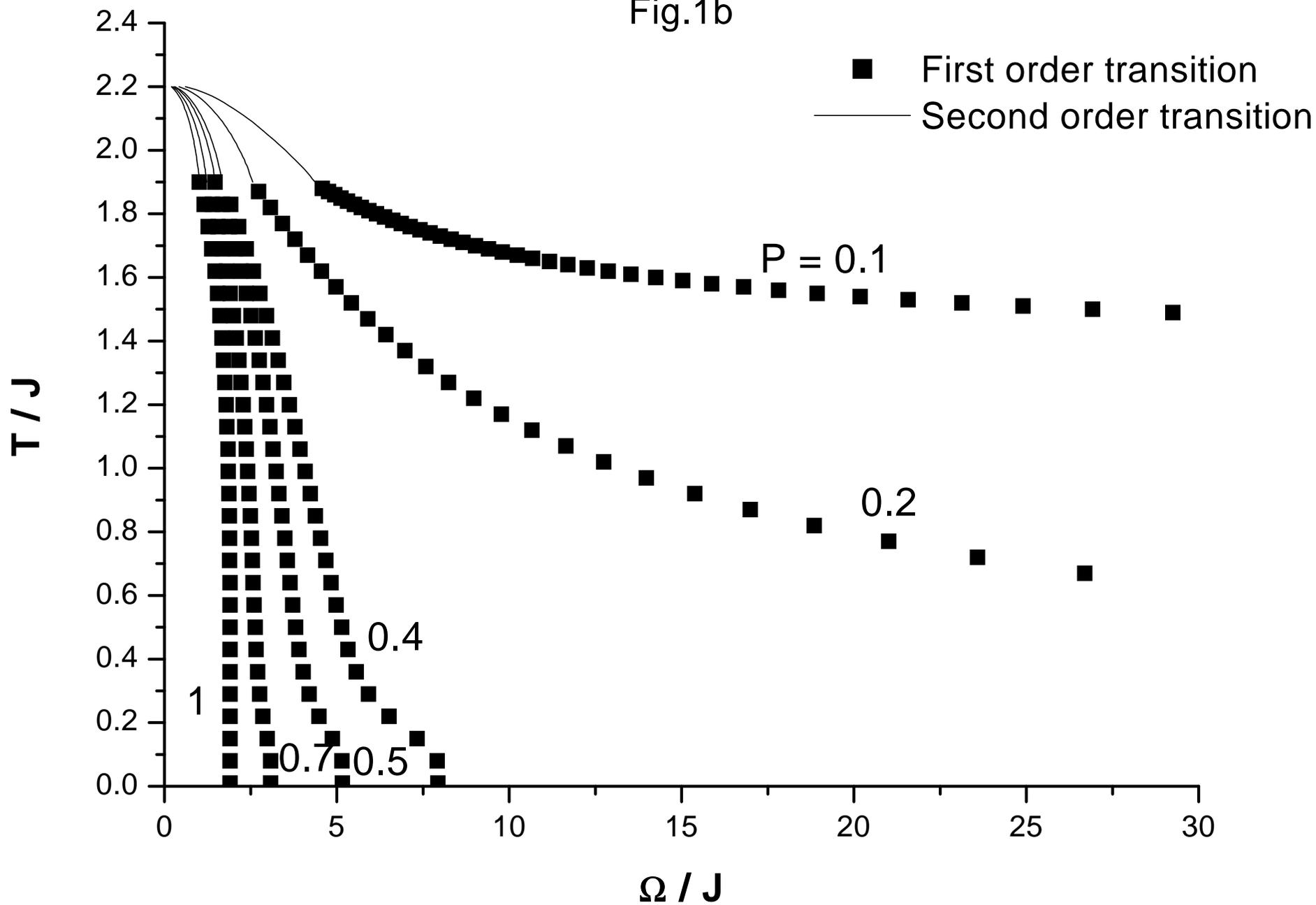
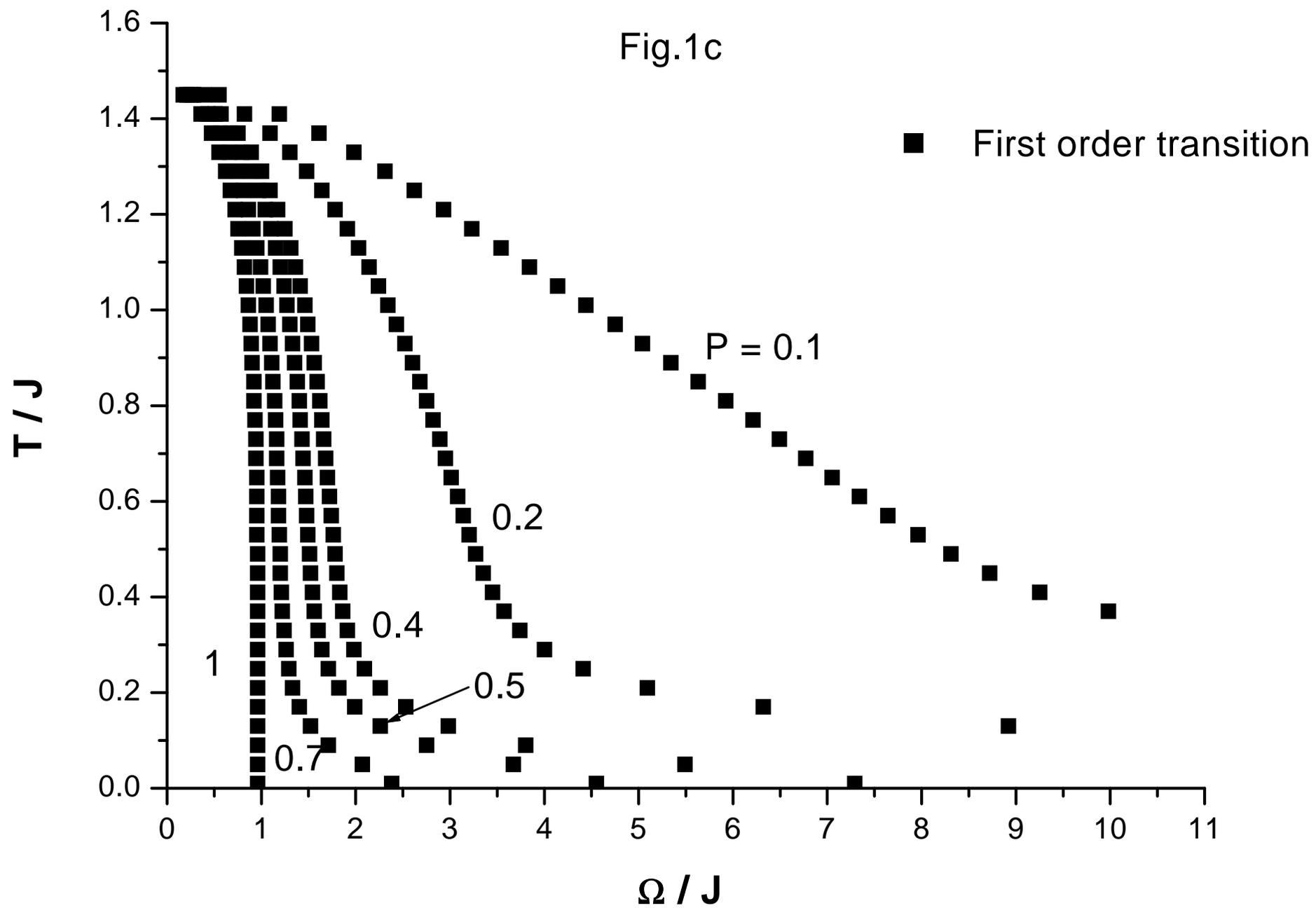
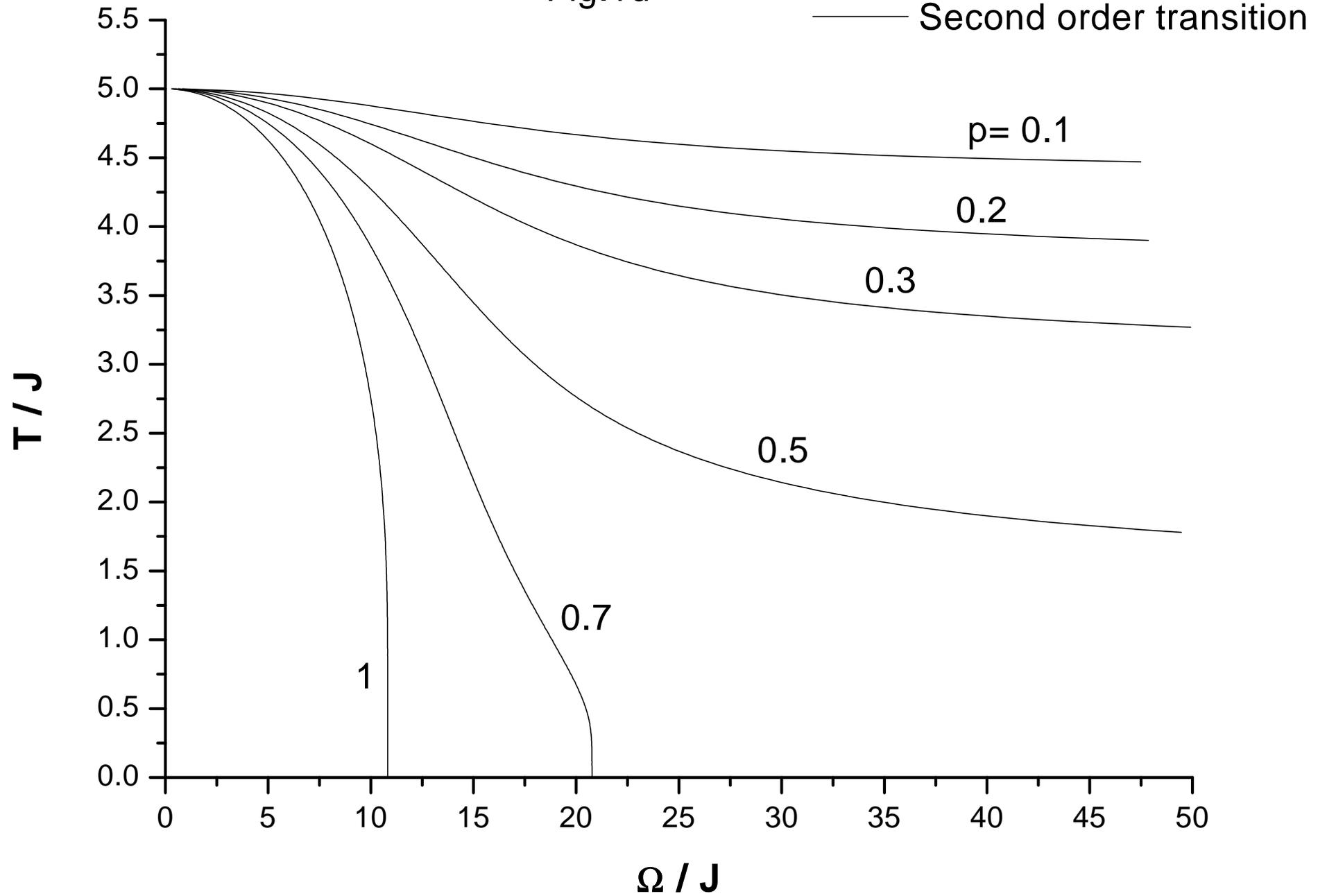
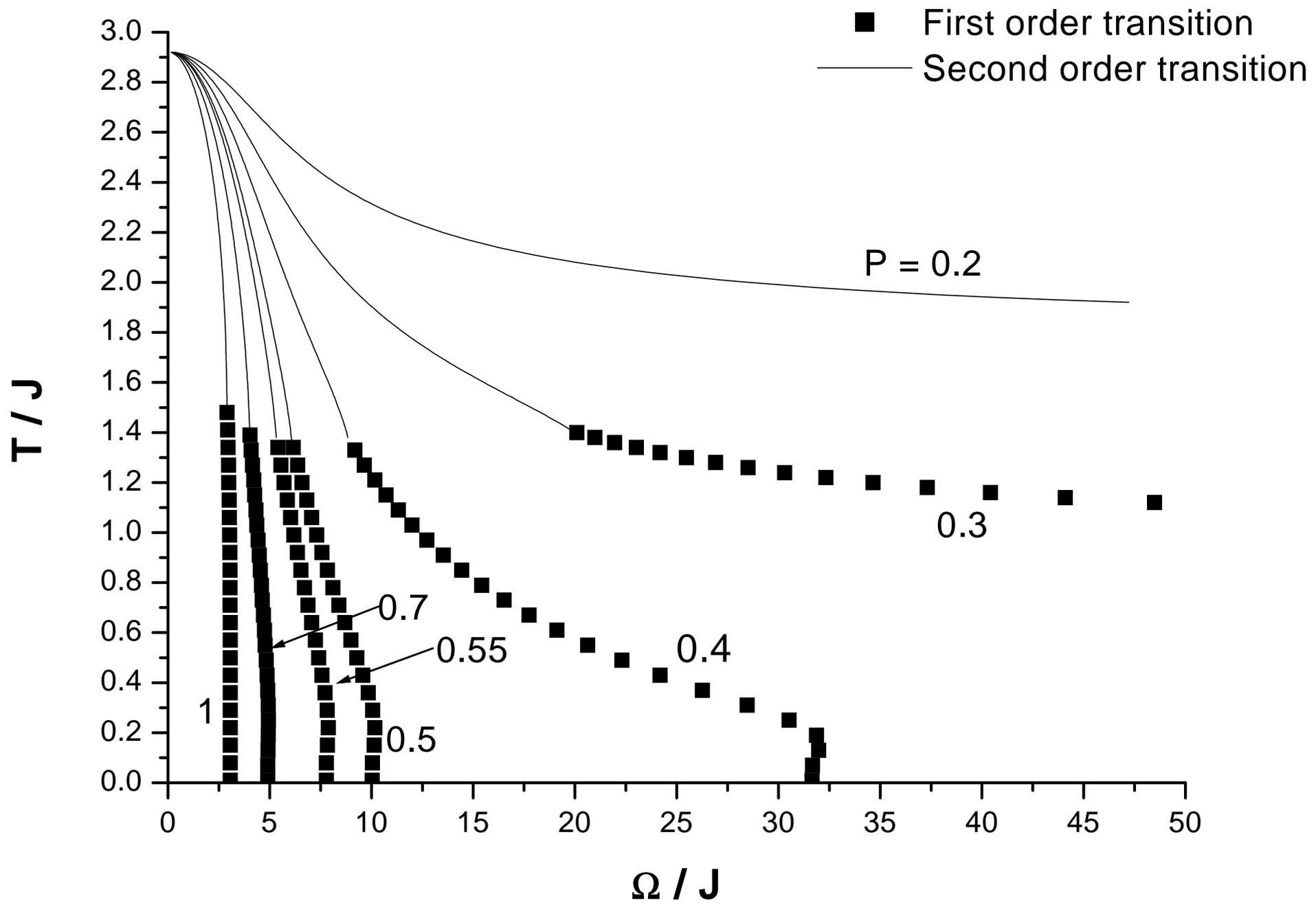
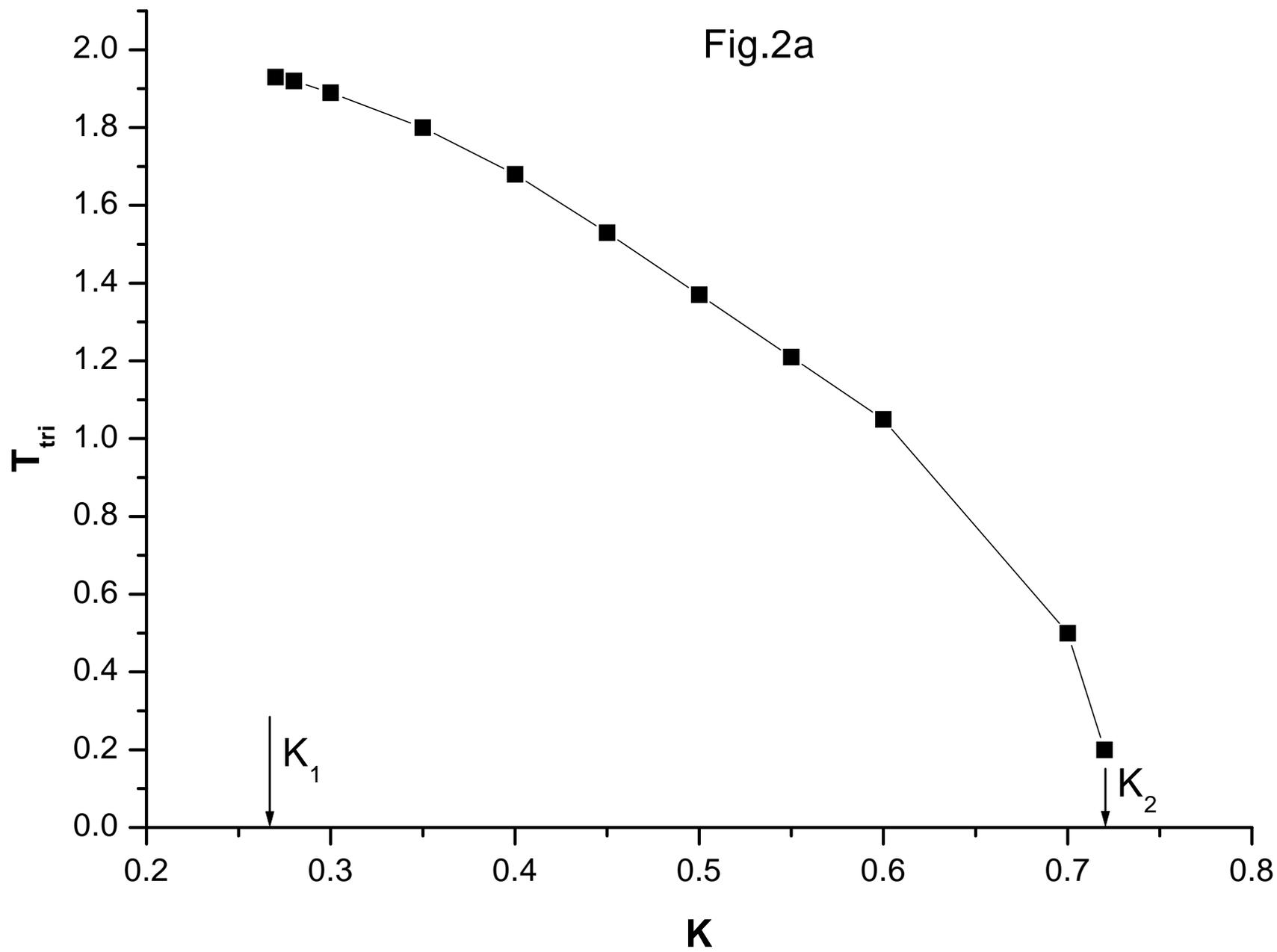
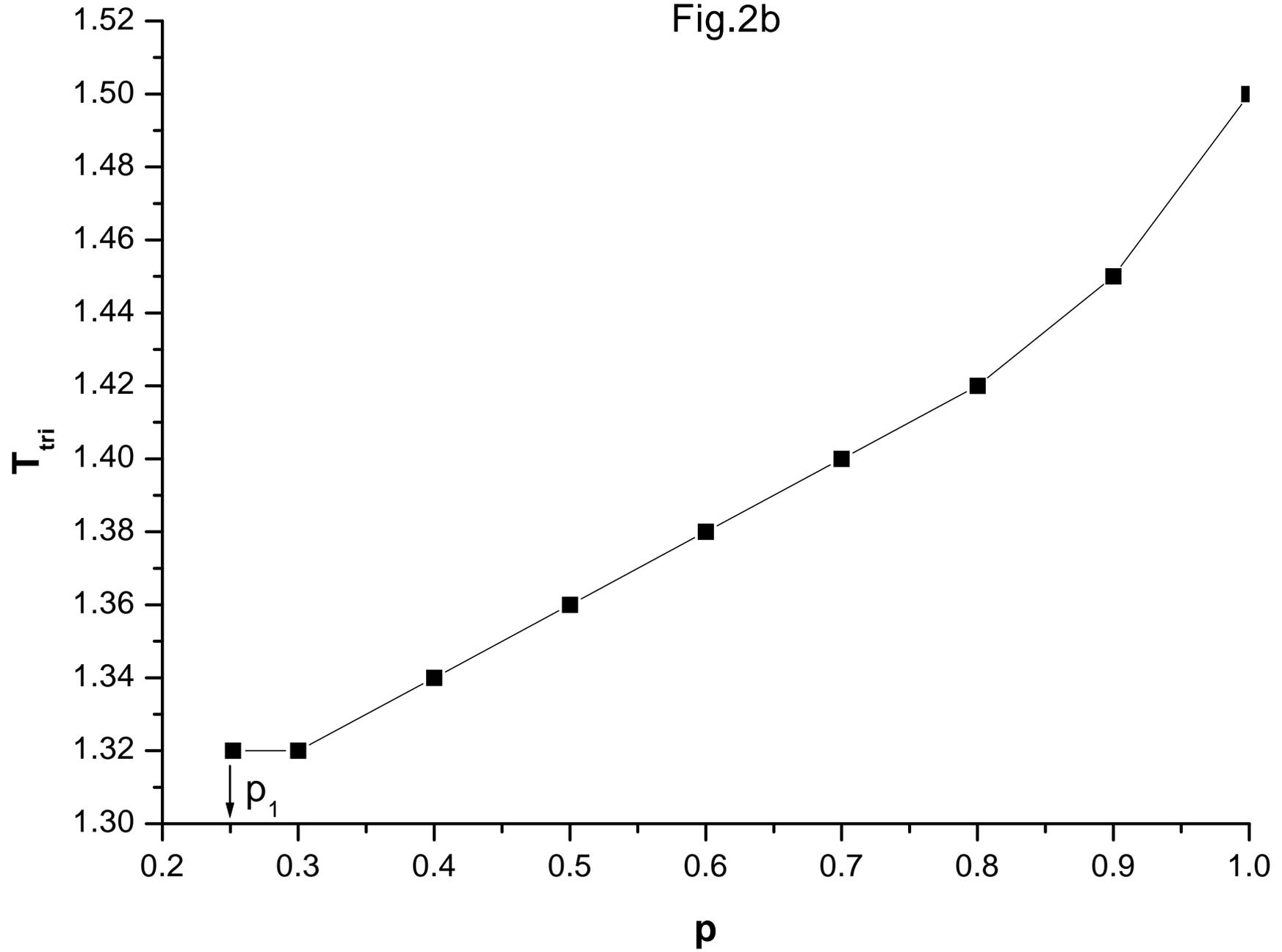
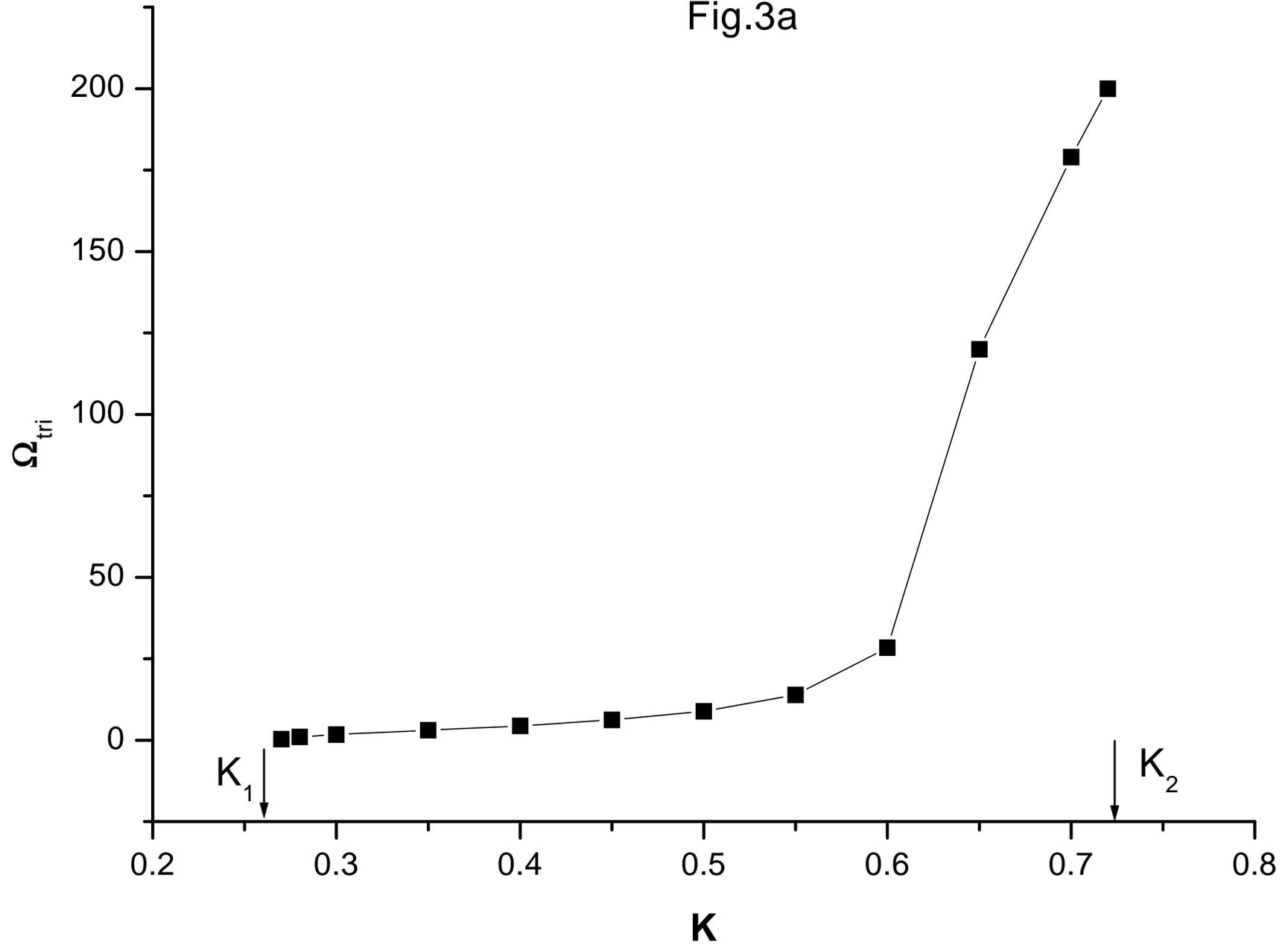
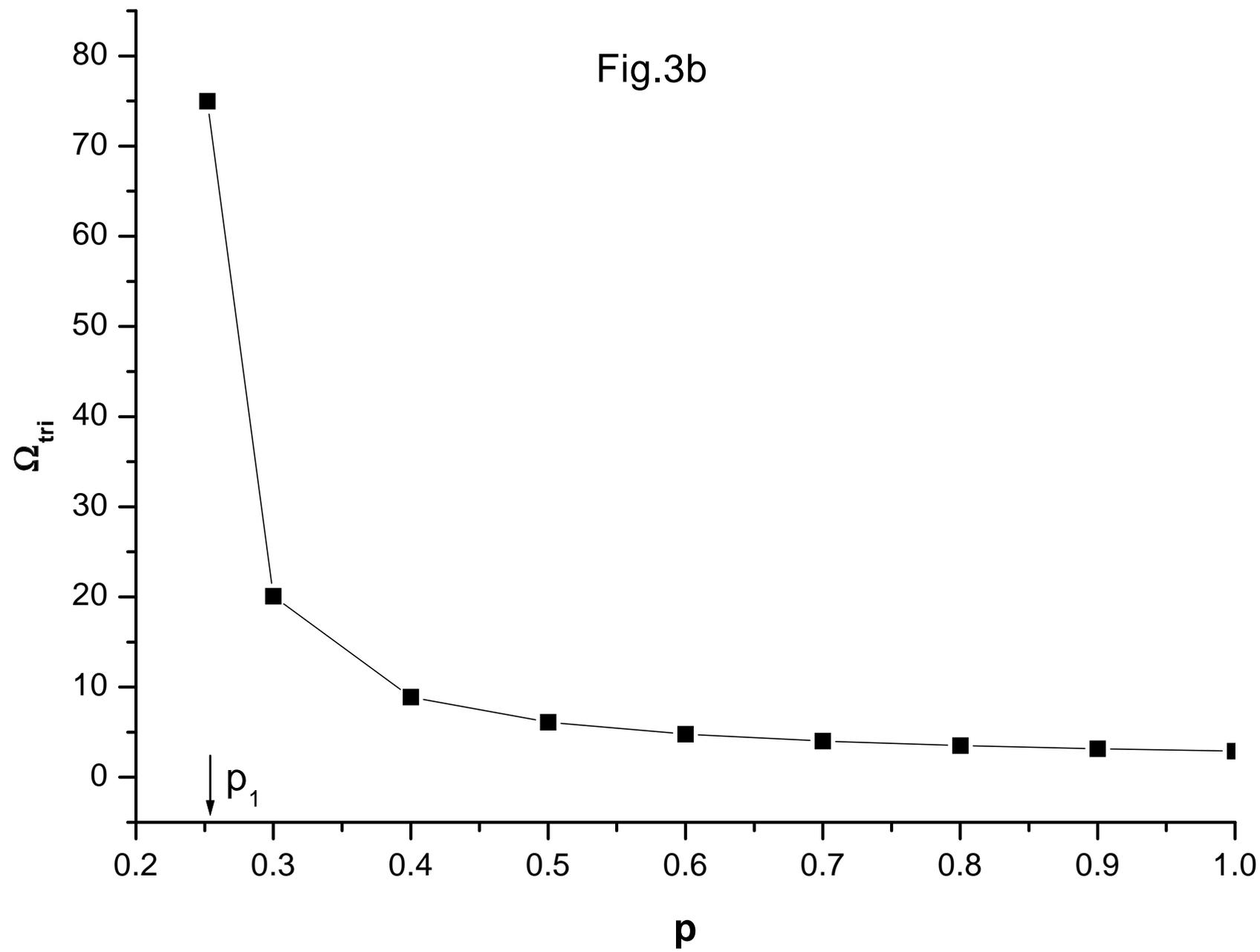
\end{document}